\documentclass[%
 reprint,
superscriptaddress,
 amsmath,amssymb,
 aip,
 jcp,
citeautoscript,
]{revtex4-2}

\usepackage[T1]{fontenc}
\usepackage{graphicx}
\usepackage{dcolumn}

\begin{document}


\title{Quantum dynamics of photophysical aggregates in conjugated polymers}

\author{Henry~J.~Kantrow}
\thanks{HJK and EGM are first co-authors of this manuscript}
\affiliation{School of Chemical and Biomolecular Engineering, Georgia Institute of Technology, Atlanta, GA~30332, United~States}

\author{Elizabeth~Guti\'errez-Meza}
\thanks{HJK and EGM are first co-authors of this manuscript}
\affiliation{School of Chemistry and Biochemistry, Georgia Institute of Technology, Atlanta, GA~30332, United~States}

\author{Hongmo~Li}
\affiliation{School of Materials Science and Engineering, Georgia Institute of Technology, Atlanta, GA~30332, United~States}

\author{Qiao~He}
\affiliation{Department of Chemistry and Centre for Processable Electronics, Imperial College London, White City Campus, London W12~0BZ, United~Kingdom}

\author{Martin~Heeney}
\affiliation{Department of Chemistry and Centre for Processable Electronics, Imperial College London, White City Campus, London W12~0BZ, United~Kingdom}
\affiliation{King Abdullah University of Science and Technology (KAUST), KAUST Solar Centre (KSC) , Thuwal 23955-6900, Saudi Arabia}

\author{Natalie~Stingelin}
\affiliation{School of Chemical and Biomolecular Engineering, Georgia Institute of Technology, Atlanta, GA~30332, United~States}
\affiliation{School of Materials Science and Engineering, Georgia Institute of Technology, Atlanta, GA~30332, United~States}

\author{Eric~R.~Bittner}
    \affiliation{Department of Physics, University of Houston, Houston, Texas 77204, United~States}

\author{Carlos~Silva-Acu\~na}
\email{carlos.silva@umontreal.ca}
\affiliation{School of Chemistry and Biochemistry, Georgia Institute of Technology, Atlanta, GA~30332, United~States}
\affiliation{School of Materials Science and Engineering, Georgia Institute of Technology, Atlanta, GA~30332, United~States}
\affiliation{Institut Courtois \& D\'epartement de Physique, Universit\'e de Montr\'eal, 1375 Avenue Th\'er\`ese-Lavoie-Roux, Montr\'eal H2V~0B3, Qu\'ebec, Canada}

\author{Hao~Li}
\email{hao.li.3@umontreal.ca}
 \affiliation{Institut Courtois \& D\'epartement de Physique, Universit\'e de Montr\'eal, 1375 Avenue Th\'er\`ese-Lavoie-Roux, Montr\'eal H2V~0B3, Qu\'ebec, Canada}

\author{F\'elix~Thouin}
\email{felix.thouin@umontreal.ca}
\affiliation{Institut Courtois \& D\'epartement de Physique, Universit\'e de Montr\'eal, 1375 Avenue Th\'er\`ese-Lavoie-Roux, Montr\'eal H2V~0B3, Qu\'ebec, Canada}

\date{21 November 2024}

\begin{abstract}
%
Photophysical aggregates are ubiquitous in many solid-state microstructures adopted by conjugated polymers, in which $\pi$ electrons interact with those in other polymer chains or those in other chromophores along the chain. These interactions fundamentally define the electronic and optical properties of the polymer film. While valuable insight can be gained from linear excitation and photoluminescence spectra, nonlinear coherent excitation spectral lineshapes provide intricate understanding on the electronic couplings that define the aggregate and their fluctuations. 
Here, we discuss the coherent two-dimensional excitation lineshape of a model hairy-rod conjugated polymer. 
At zero population waiting time, we find a $\pi/2$ phase shift between the 0--0 and 0--1 vibronic peaks in the real and imaginary components of the complex coherent spectrum, as well as a dynamic phase rotation with population waiting time over timescales that are longer than the optical dephasing time. 
We conjecture that these are markers of relaxation of the photophysical aggregate down the tight manifold of the exciton band. 
These results highlight the potential for coherent spectroscopy 
via analysis of the complex spectral lineshape 
to become a key tool to develop structure-property relationships in complex functional materials.
\end{abstract}

\maketitle

\section{Introduction}

Conjugated polymers are complex photophysical systems where vibronic 
    nearest-neighbor couplings lead to exciton dispersion in a disordered energy landscape. Polymers can adapt a diversity of solid-state microstructures that depend both on materials parameters, such as the chemical structure and molecular weight, and processing conditions. This diversity of microstructures presents a variety of ways in which $\pi$ electrons in a backbone chromophore interact with other chromophores, whether in a neighboring chain or within the same chain~\cite{reid_influence_2012}. Photophysical aggregates arise from the ensemble of such $\pi$-electron correlations; when $\pi$--$\pi^{*}$ transition dipole moments couple in a co-facial manner, so-called H aggregates arise. Conversely, when they interact in a head-to-tail fashion, one considers J aggregates. Consequently, H aggregates mostly reflect inter-chain interactions, while J aggregates those within a chain. In many conjugated polymer solid-state microstructures, aggregates have hybrid HJ character\cite{yamagata_interplay_2012,spano_h-_2014}, producing an electonically rich and complex disordered energy landscape. Examples of such interchromophore interactions are resonance Coulomb correlations, and the details of these fundamentally determine the optical properties.  

Within this picture, primary photoexcitations in neat $\pi$-conjugated polymers in the solid state are Frenkel excitons~\cite{pope_electronic_1999}, with the absorption and photoluminescence (PL) spectral lineshapes modeled quantitatively using a photophysical aggregate model that accounts for excitonic inter-chromophore coupling, electron-vibrational coupling, and energetic disorder~\cite{spano_modeling_2005,spano_erratum_2007,clark_role_2007,clark_determining_2009,spano_determining_2009,yamagata_interplay_2012,paquin_two-dimensional_2013,spano_h-_2014}.
Within this model, the spectral signatures depend heavily on the polymer microstructure, and more locally, chain conformations~\cite{paquin_two-dimensional_2013}.
Effective excitonic couplings and the magnitude of energetic disorder can all be extracted from the linear absorption spectrum~\cite{spano_modeling_2005,spano_erratum_2007,clark_role_2007,clark_determining_2009}.
Combined with analysis of the PL spectrum, one can infer the two-dimensional average coherence function and the disorder spatial correlation~\cite{spano_determining_2009,paquin_two-dimensional_2013}.

While such analysis of the linear spectra have provided important insights into the photophysics of neat polymers and blends, they are insensitive to a host of phenomena in these complex materials. 
One challenge is that the linewidths of most conjugated polymers are quite broad due to the complex disorder landscape. 
This line broadening masks the contribution of a rich fine structure, such as low energy vibronic replicas and pairs of aggregate states.
Beyond broadening, these spectroscopic observables are insensitive to the coherent and incoherent channels between vibronic states which mediate excitonic relaxation. 
Linear optical probes are also blind to many-body effects and higher-lying-state manifolds central to processes that control optoelectronic device performance, including bimolecular annihilation, singlet fission or triplet fusion.

Non-linear coherent spectroscopic techniques like multidimensional coherent electronic spectroscopy have shown they can directly probe these phenomena in materials systems with relatively narrow linewidths, including atomic vapors~\cite{tian_femtosecond_2003,dai_two-dimensional_2010}, GaAs quantum wells~\cite{zhang_polarization-dependent_2007,turner_coherent_2010,turner_persistent_2012}, transition metal dichalcogenides~\cite{moody_intrinsic_2015,czech_measurement_2015,hao_neutral_2017} and even dynamically disordered systems such as 2D and 3D metal-halide perovskites~\cite{thouin_stable_2018,thouin_enhanced_2019, camargo_dark_2020}.
By analyzing coherent lineshapes in these experiments, a host of subtle effects were probed like excitation induced shift and dephasing~\cite{shacklette_role_2002}, multi-excitonic binding~\cite{turner_coherent_2010}, dark states~\cite{tollerud_revealing_2016}, and independent measurement of the homogeneous and inhomogeneous linewidths~\cite{moody_exciton-exciton_2011}.

In organic systems, initial multidimensional spectroscopic measurements have focused on the dynamics of spectral correlations, such as excited state absorption, bleaching~\cite{de_sio_tracking_2016} and vibrational coherences~\cite{camargo_resolving_2017,song_vibrational_2014,song_separation_2015}.
Other investigations have observed multiexcitonic states~\cite{gutierrez-meza_frenkel_2021, zheng_unveiling_2024} using higher-order measurements.
Despite the availability of broadband sources covering most of the absorption spectra of organic chromophores and extensive theoretical analysis, few reports leverage the richness of the coherent lineshapes in multidimensional coherent electronic spectroscopy.

In this communication, we highlight the potential and need for coherent lineshape analysis in $\pi$-conjugated polymers. 
We focus on a prototypical polymer poly(2,5-bis(3-hexadecylthiophene-2-yl)- thieno[3,2-\textit{b}]thiophene) (PBTTT, Fig.~\ref{fig:AbsPL}a), which has a hairy-rod structure due to the fused rings in its backbone and long sidechain substitutions~\cite{mcculloch_liquid-crystalline_2006,snyder_classification_2015,delongchamp_molecular_2008}. 
These two effects can induce liquid-crystalline-like behavior in PBTTT, producing a solid-state structure that often is highly structurally dynamic~\cite{poelking_characterization_2013,wunderlich_difference_1996,wunderlich_mcromolecular_1976,zhang_inplane_2010}.
We first analyze the absorption and emission spectra of PBTTT using the framework of the weakly-coupled hybrid HJ-aggregate model.
We then attempt to interpret the rephasing multidimensional coherent spectrum and its evolution with interpulse delay using the same framework.  These 2D spectra are able to distinguish between homogeneous and inhomogeneous broadening, potentially revealing additional features hidden under the broad linear linewidths.

\section{Results and Discussion}

\subsection{Analysis of linear lineshapes}

\begin{figure}
\begin{center}
\includegraphics[width=8cm]{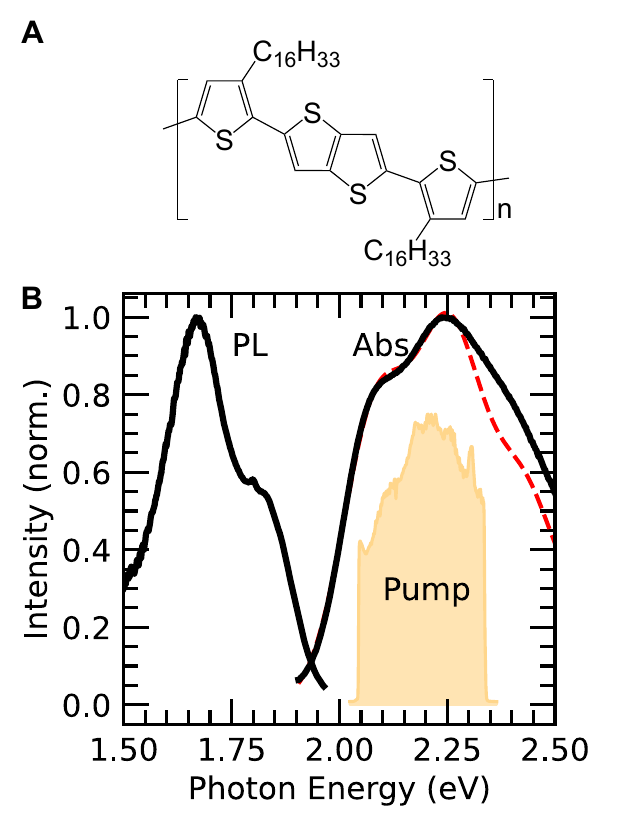}
\caption{(a) Chemical structure of PBTTT. (b) Normalized photoluminescence and absorbance spectra at 5\,K (labeled black lines) and fit of the absorption spectrum (red dashed line) to a modified Franck-Condon progression (Eq.~\ref{eq:FC}) with $W=31$\,meV, $\hbar \Omega_{\mathrm{vib}} = 170 \pm 10$\,meV, $S = 1$, and $\sigma = 77$\,meV.}
\label{fig:AbsPL}
\end{center}
\end{figure}

We first consider the lineshapes of the PBTTT linear absorption and photoluminescence (PL) spectra because they contain intricate information excitonic intra- and intermolecular coupling and energetic disorder~\cite{spano_h-_2014}. Fig.~\ref{fig:AbsPL} displays the linear absorption and photoluminescence spectra of a PBTTT film measured at 5\,K. These spectra display lineshapes consistent with the hybrid HJ-aggregate model by Yamagata and Spano~\cite{yamagata_interplay_2012}, in which the two-dimensional spatial coherence function of the vibronic Frenkel exciton depends on the interplay of intrachain and interchain resonance electronic Coulomb coupling~\cite{paquin_two-dimensional_2013}. 

The dominant interaction reflected in the spectra in Fig.~\ref{fig:AbsPL}b is the interchain (H-like) coupling: the 0--0 peak is suppressed with respect to the rest of the vibronic progression.
We thus model the absorption spectrum using the weakly coupled H-aggregate model: a modified Franck-Condon progression that perturbatively considers interchain excitonic coupling $J$~\cite{spano_modeling_2005,clark_determining_2009}.  The model includes a single effective normal mode with energy $\hbar \Omega_{\mathrm{vib}}$, an effective Huang-Rhys parameter $S$, and a Gaussian lineshape for each vibronic peak with standard deviation $\sigma_{\mathrm{abs}}$~\cite{yamagata_interplay_2012}. The coupling term $J$ is reflected in the free-exciton bandwidth $W=4J$. 
The linear absorption lineshape then reads as 
 \begin{equation} \label{eq:FC}
     \begin{split}
         A(\omega)  \propto & \sum_{m=0}  \frac{e^{-S}S^m}{m! \sqrt{2\pi}\sigma_{\mathrm{abs}}}  \left( 1 - \frac{W e^{-S}}{2 \hbar \Omega_{\mathrm{vib}} }  \sum_{\nu \neq m} \frac{S^\nu }{\nu!(\nu-m)} \right) \\
 & \times \exp \left[-\frac{(\hbar \omega - \hbar \omega_{\mathrm{0}} - m\hbar \Omega_{\mathrm{vib}})^2}{2 \sigma_{\mathrm{abs}}^2}\right],
     \end{split}
 \end{equation}
where $E_{0}$ is the origin energy of the vibronic progression, and $\nu$ is a vibrational quantum number~\cite{spano_modeling_2005}. 

The modeled absorption spectrum (Fig.~\ref{fig:AbsPL}b) deviates from the experimental one at higher energies, where non-aggregate absorption dominates~\cite{clark_role_2007,clark_determining_2009}. We find that the best fit yields $W = 31$\,meV, $\hbar \omega_{\mathrm{vib}} = 170  \pm 10$\,meV, and $\sigma_{\mathrm{abs}} = 77$\,meV.  The value of $W << \hbar \omega_{\mathrm{vib}}$ is consistent with the weakly coupled H-aggregate model~\cite{spano_modeling_2005}.

We now turn to the PL lineshape, which reflects the details of the disorder width and spatial correlation parameters. While the 0--0/0--1 absorbance ratio depends dominantly on $W$, the scaling of the 0--0/0--1 PL ratio also depends on disorder parameters, including the exciton coherence length~\cite{spano_determining_2009}. In general, energetic disorder relaxes the symmetry selection rules that suppress 0--0 emission in a weakly coupled H-aggregate. In the limit of a spatially uncorrelated disorder landscape, the 0--0 intensity relative to the rest of the vibronic progression grows with the disorder parameter ($\sigma_{\mathrm{em}}$). 
However, the disorder landscape can adopt spatial correlation due to the conformation of a chain within a single aggregate. The disorder within an aggregate is quantified by the spatial correlation length ($\ell_0$). Defining the spatial correlation parameter $\beta \equiv \exp(-1/\ell_0)$, the ratio of 0--0 and 0--1 vibronic intensities in PL depends on $\sigma_{\mathrm{em}}$, $\beta$, and $W$ as~\cite{paquin_two-dimensional_2013}
\begin{equation} \label{eq:PL_ratio}
    \frac{I_{0-0}}{I_{0-1}} \sim \frac{(1-\beta)\sigma_{\mathrm{em}}^2}{(1+\beta)W^2},
\end{equation}
which is valid in the limit of weak energetic disorder compared to the free-exciton bandwidth~\cite{spano_determining_2009}.  

In Fig.~\ref{fig:AbsPL}b, we observe that the PL spectrum displays a Huang-Rhys parameter $S = 2I_{0-2}/I_{0-1} \approx 1$, which is consistent with that used in the fit of the absorption spectrum (Eq.~\ref{eq:FC}). 
However, the PL $I_{0-0}/I_{0-1}$ is roughly 0.55, which is lower than the corresponding ratio in the absorption spectrum. 
All other parameters taken from the fit of the absorption spectrum, we estimate that $\beta \sim 0.78$ by eq.~\ref{eq:PL_ratio} and therefore $\ell_0 \sim 4$.

This $I_{0-0}/I_{0-1}$ PL ratio and spatial correlation parameter ($\beta$) is comparable to that found in the flexible-chain poly(3-hexylthiophene) (P3HT)~\cite{paquin_two-dimensional_2013}. 
From this linear analysis, we can establish that excitons in PBTTT experience a comparably disordered landscape with comparable correlation to that of P3HT, despite the more rigid PBTTT backbone. This is unsurprising, as the correlated disorder landscape in P3HT was previously shown to be weakly dependent on the $M_w$ and resulting degree of backbone torsional order~\cite{paquin_two-dimensional_2013}.  This highlights the limitations of linear spectroscopy to unravel the complex disorder landscape in conjugated polymers. 

\subsection{Coherent nonlinear spectroscopy}

\begin{figure*}
\begin{center}
\includegraphics[width=12.9cm]{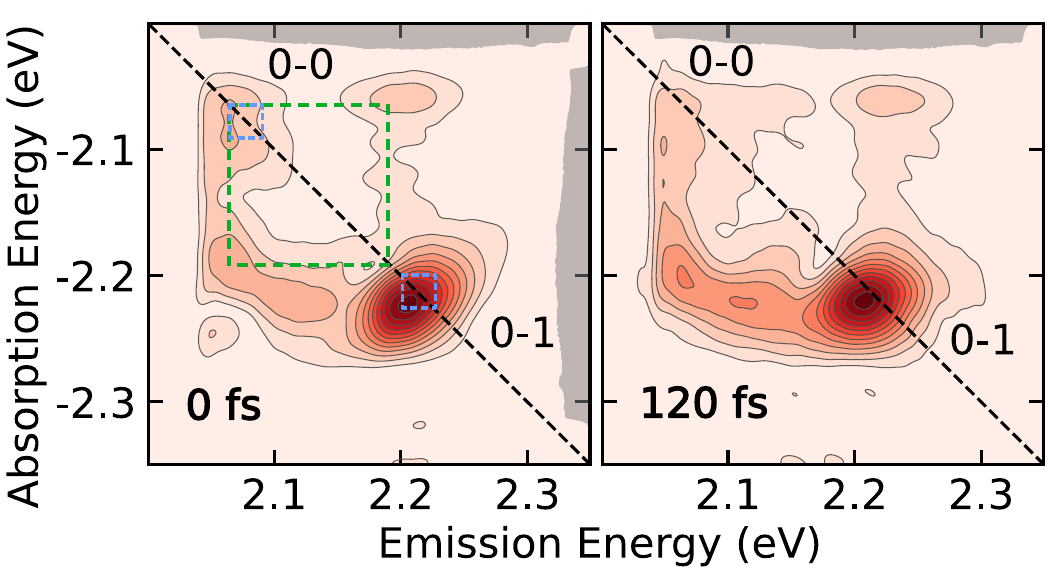}
\caption{Norm of the 2D coherent excitation spectrum of PBTTT measured at 5\,K, with a rephasing pulse sequence and population time of (left) 0 and (right) 120\,fs. Dashed squares indicate cross peaks and their related features on the diagonal with energies of 24\,meV (blue) and 125\,meV (green). The shaded gray border denotes the spectrum of the femtosecond pulse train used for the experiments, which partially covers the 0--0 transition and fully covers the 0--1 vibronic replica.}
\label{fig:Abs2D}
\end{center}
\end{figure*}

We now turn to multidimensional spectroscopy to obtain additional insights into photophysics that would otherwise be buried in its broadened linear spectra. The 2D coherent lineshape of flexible-chain $\pi$-conjugated polymers typically reflects strong inhomogeneous broadening due to static disorder. This presents itself as elongation along the $\hbar\omega_{\mathrm{abs}} = \hbar\omega_{\mathrm{em}}$ diagonal, where the cross peaks between vibronic transitions are thus largely obscured by this broadening~\cite{song_vibrational_2014,song_separation_2015,gregoire_excitonic_2017,gregoire_incoherent_2017}. 
However, the scenario in the hairy-rod polymer PBTTT is entirely different. We first focus on the absolute rephasing signal of PBTTT at 5\,K and zero population time (Fig.~\ref{fig:Abs2D}, left panel).
This spectrum shows a very well defined structure in the form of diagonal 0--0 and 0--1 features at 2.06 and 2.21\,eV, respectively.
Unlike the case in most flexible chain polymers, these vibronic peaks are only weakly elongated along the diagonal.


We will discuss extensively the detailed relationship between conjugated polymer microstructure and the disordered landscape in a separate publication, in which we will compare similar observations with those in flexible-chain polymer systems. Here, we simply note that the disordered energy landscape is in a dynamic regime in order to discuss the lineshape evolution with population waiting time.  
Upon fitting the 0--1 transition of the early-time spectrum in Fig.~\ref{fig:Abs2D}, we find a homogeneous linewidth of $37 \pm 2$\,meV and an inhomogeneous linewidth of $5 \pm 3$\,meV (Supplementary Fig.~S2)~\cite{siemens_resonance_2010}. 
This means that the magnitude of static electronic disorder is comparable to that of dynamic disorder in PBTTT. 
Here, we note that the coherent \emph{rephasing} linewidth ($\sim45$\,meV) is considerably more narrow than the linear \emph{total} linewidths ($\sim70$\,meV). 
This difference may indicate that additional line-broadening mechanisms triggered by photoexcitation occur over longer time scales than the femtosecond dynamics of our experiment.
We also highlight that the spectra observed using this four-wave mixing implementation of multidimensional coherent spectroscopy are considerably sparser than those observed in PBTTT using population-detected methods~\cite{gutierrez-meza_frenkel_2021}. 
In population detection methods, asymmetries between the effect of the population probe (e.g.\ fluorescence) on the singly and doubly excited state manifolds yield additional coherent pathways that contribute to the detected signal~\cite{perdomo-ortiz_conformation_2012}.
For example, in a system composed of a common ground state, two nondegenerate singly excited states, and a doubly occupied state at the sum of their energies, additional pathways appear as cross-peaks that would otherwise be invisible in a four-wave mixing experiment.
As polymer aggregates consist of many closely spaced vibronic states, it thus follows that the population-detected multidimensional spectrum of PBTTT is more congested than its four-wave mixing equivalent.

In addition to these diagonal peaks, we identify a rich cross-peak structure highlighting the many coherent pathways coupling the vibronic states.
As expected from the linear absorption spectrum, cross peaks are observed at ($\hbar\omega_{\mathrm{abs}} ,\hbar\omega_{\mathrm{em}}$) $\sim$ ($\pm$2.06, $\mp$2.24) (Fig.~\ref{fig:Abs2D}, left panel).
These features are expected for the vibronic progression observed in PBTTT, as the 0--0 and 0--1 features are spectrally correlated through their common ground state.
In comparison, the lower energy cross peak at $(\hbar\omega_{\mathrm{abs}} ,\hbar\omega_{\mathrm{em}}) \sim (\pm2.06, \mp2.2)$ connects the 0--0 state to a feature at 2.2\,eV which is absent on the diagonal (green box in Fig.~\ref{fig:Abs2D}).
This indicates the presence of a dark state observable in the multidimensional spectrum through its coherent coupling with bright 0--0 state\cite{tollerud_revealing_2016}.

Low energy cross-peaks around both 0--0 and 0--1 diagonal features also arise (blue dashed squares in Fig.~\ref{fig:Abs2D}), shifted by 24\,meV from the diagonal. These represent coupling of states close in energy to the main 0--0 and 0--1 vibronic transitions.
For both these cross-peak complexes, features at lower emission energy have greater amplitude, indicating an incoherent decay channel within the fine structure of each diagonal peak.
As the electronic states are coupled to a host of vibrational modes beyond those responsible for the main progression, we attribute this fine structure to underlying vibronic progressions within each manifold.


We next consider the temporal evolution of the 2D lineshapes at a long population time of 120\,fs (Fig.~\ref{fig:Abs2D}, right panel). The norm of the 2D spectra at intermediate $t_{\mathrm{pop}}$ are provided in Supplementary Fig.~S3.
During the probed population time delay of 120\,fs, diagonal and cross-peak features identified at 0\,fs broaden, while their relative ratio remain similar.
We interpret this as a signature of spectral diffusion, where decay channels coupling closely spaced energy levels within a vibronic band redistribute the populations and coherences~\cite{singh_localization_2017,roberts_characterization_2006}. 
This leads to the elongation of the 0--1 feature along the diagonal and loss of the fine structure observed in the 0--0 diagonal features and their cross-peaks.

We also note the growth of the cross peak at ($\hbar\omega_{\mathrm{abs}},\hbar\omega_{\mathrm{em}}$) $\sim$(--2.2, 2.06), assigned to the coherence between a bright state in the 0--0 manifold and a dark state close to the 0--1 manifold. 
We interpret the increased amplitude of this cross-peak as a signature of the decay of population in the 0--1 manifold into this mixed bright-dark coherence.
The continued absence of the associated diagonal peak at ($\hbar\omega_{\mathrm{abs}} ,\hbar\omega_{\mathrm{em}}$) $\sim$(--2.17, 2.17) further supports this assignment, as it would arise from the transfer of population in the 0--1 manifold into a dark population insensitive to further light-matter interactions.
We also note the growth of a diagonal feature at $(\hbar\omega_{\mathrm{abs}} ,\hbar\omega_{\mathrm{em}}$) $\sim$ (--2.15, 2.15).
Due to its separation from the bottom of the 0--0 manifold, this feature could arise from the population of a vibrational manifold with a $\sim 90$\,meV ($725$\,cm$^{-1}$) spacing.
This energy corresponds to a low-intensity mode previously observed in resonance Raman spectrum of PBTTT~\cite{furukawa_raman_2016}.

\begin{figure*}
\begin{center}
 \includegraphics[width=17.4cm]{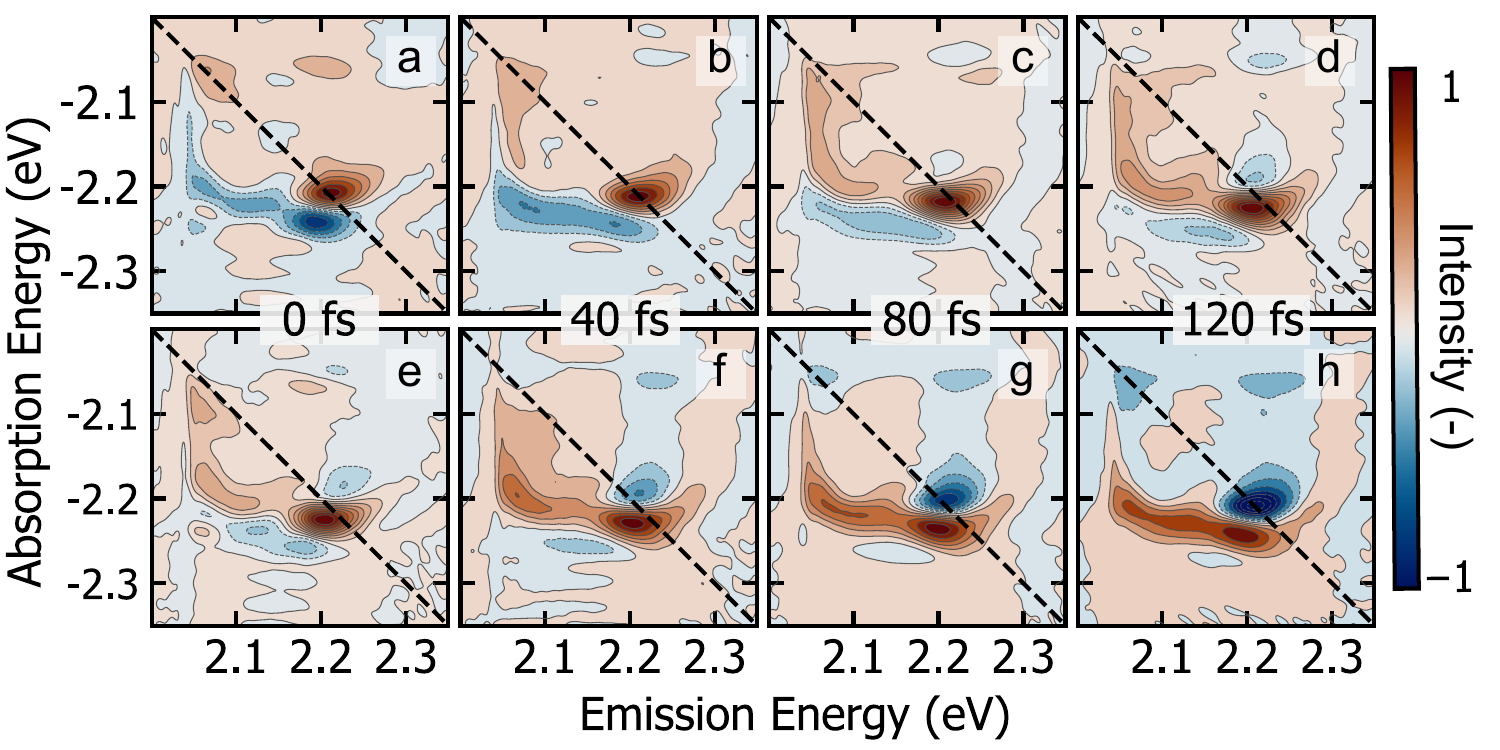}
\caption{(a-d) Real and (e-h) imaginary part of the rephasing coherent 2D spectra of a PBTTT film, measured at 5\,K, as a function of population waiting time $t_{\mathrm{pop}}$ labeled on the spectra.}
\label{fig:complex}
\end{center}
\end{figure*}

We now turn to the description of the coherent lineshape and focus on the complex (real and imaginary) parts of the rephasing spectra as a function of population time (Fig.~\ref{fig:complex}, see Supplementary Fig.~S4 for the full $t_{\mathrm{pop}}$ dependence).
At zero population time, the real component of the diagonal 0--0 feature is symmetric around the diagonal.
This so-called absorptive lineshape is characteristic of the real part of the rephasing lineshape of an inhomogeneously broadened transition~\cite{li_optical_2023}.
The 0--0 lineshape evolves with population time towards a clearly anti-symmetric (dispersive) lineshape around the diagonal by a population time of 120\,fs. 
The imaginary component of the 0--0 diagonal feature features the opposite trend, converting from dispersive to absorptive along the population time evolution. 
Likewise, the real part of the 0--1 transition features a dispersive lineshape at zero population time which later becomes absorptive. We underline that these phase rotation dynamics persist on a timescale much longer than the dephasing time, and that they persist over other population evolution mechanisms such as spectral diffusion. 

To understand the cause of this phase rotation, we note that the spectral lineshape primarily relies on the ensemble average of the corresponding response function, in which the cumulant or Magnus expansion is often applied~\cite{mukamel_principles_1995}. Such evaluation involves the product of exponential functions at different times $t$ in the form of $\exp\left[\pm \frac{i}{\hbar}\int_0^t H_0(\tau) \rm d\tau\right]$. In the simplest case of a cumulant expansion to the second order, which is exact for models of Gaussian statistics, one obtains
\begin{align}
    \left<\exp\left[\pm \frac{i}{\hbar}\int_0^t H_0(\tau) \rm d\tau\right]\right> = e^{\pm (i/\hbar)g_1(t)} e^{-g_2(t)/(2\hbar^2)},
\end{align}
in which $g_1(t)$ and, in general with two time limits, $g_2(t,t^{\prime})$, 
correspond to the first and second moments of the exponent, respectively. 
Therefore, $g_1(t)$, namely, the \emph{line shift function}, introduces a frequency shift in a linear spectrum and a phase shift in the homogeneous line shape in a 2D spectrum.
On the other hand, $g_2(t,t^{\prime})$ only contributes to the line broadening. 

The observed phase evolution in the homogeneous line shape of PBTTT suggests a nontrivial first-order cumulant for the exponential term involving population time $t_2$. 
In other words, the average of the time evolution of the density operator over $t_2$ in the interaction picture is nontrivial. 
This could arise from either the evolving system Hamiltonian, e.g., nonstationary excitation induced dephasing, \cite{li_optical_2023-1} or unevenly weighted Liouville space pathways in the excited- or ground-state manifold. 
The latter is much more likely in the PBTTT aggregates, due to the low fluence used here (10 nJ cm$^{-2}$) and the typical weak fluence dependence of the coherent lineshape in organic systems~\cite{zheng_unveiling_2024}.

We interpret the opposite phase evolution of the 0--0 and 0--1 features as a signature of relaxation down a tight manifold of the exciton band. 
In the H-aggregate with positive interchain coupling ($J > 0$), the 0--0 origin is suppressed in absorption with respect to the rest of the vibronic progression, carrying the signatures of excitons in the photophysical 
aggregate~\cite{clark_role_2007,spano_h-_2014,bittner_theory_2007}. 
The 0--1 transition, along with the rest of the vibronic progression, instead reflects the behavior of the molecular excitation independent of the aggregate. This is manifested, for example, in population as measured via time-resolved PL~\cite{paquin_two-dimensional_2013}. 
Taking the stimulated emission and excited-state absorption rephasing spectra as an example, the unevenly distributed 0--0$^{\prime}$/0--1$^{\prime}$ absorption leaves the residue of $e^{i\omega_{ee^{\prime}}t_2}$ in the response function, where $e$ and $e^{\prime}$ denote fine energy level structure in the same excitation manifold. 
In the 0--1 manifold, the complex components shift about $\pi/2$ (from absorptive to dispersive or vice versa) over 100 fs, which indicates $\hbar\omega_{ee^{\prime}}\approx 10~\rm{meV}$, corresponding to the fine vibronic structure.


Moreover, the dynamics of the 0--0 and 0--1 lineshape likely reflect ultrafast evolution of the PBTTT aggregate. 
Time-resolved PL measurements in P3HT have shown that the 0--0/0--1 ratio relaxes on picosecond timescales, much faster than the excited-state lifetime, due to relaxation in the excited state of torsional normal modes along the polymer backbone. 
As the exciton evolves from a low to high symmetry state, the relative PL 0--0 intensity decreases dynamically~\cite{parkinson_role_2010}. 
However, the exciton becomes more delocalized along the chain as a result of this process, decreasing the interchain excitonic coupling $J$ ~\cite{paquin_two-dimensional_2013}. 
This intrachain delocalization would instead increase the 0--0/0--1 ratio. 

The evolution of the cross peak between the 0--0 and 0--1 transitions at ($\hbar\omega_{\mathrm{abs}} ,\hbar\omega_{\mathrm{em}}$) $\sim$ (2.2, 2.06) eV likely arises from a dynamic increase in the exciton coherence length along the chain. 
The torsional relaxation dynamics process occurs by the normal modes with $\sim13$\,ps period~\cite{parkinson_role_2010}, which is too slow for a $\sim 120$\,fs time window. 
This indicates that the exciton delocalization is not limited by torsional relaxation dynamics. 
Instead, the rapid timescale is consistent with ultrafast depolarization dynamics observed in this class of materials, which have been interpreted as electronic relaxation~\cite{song_separation_2015}. 

This rich spectral evolution contains a wealth of information that can only be fully deciphered via advanced quantum dynamical modeling, propagating the Holstein Hamiltonian to calculate nonlinear spectral response functions. Such analysis can test hypotheses that map specific quantum dynamics to time-resolved features in the complex coherent spectral lineshape. This is a substantial undertaking, hence we anticipate that our work will stimulate future activities from the broad polymer photophysics community. 

\section{Conclusions}
By comparing the lineshapes in the linear absorption and PL spectra of PBTTT to its rephasing multidimensional lineshapes, we highlighted the potential of nonlinear coherent spectroscopy to overcome shortcomings of linear spectroscopic techniques in dissecting the intricate photophysics of conjugated polymers.

Overall, we showed the ability of the rephasing non-linear coherent experiment to resolve fine structure otherwise buried under inhomogeneous or time-averaged broadening in photophysical aggregates of conjugated polymers.
The evolution of the absolute rephasing lineshapes with population time allows for direct observation of dynamic disorder and the timescales of spectral diffusion.
The presence of asymmetrical cross-peaks also isolates the energy of the vibrational modes involved in the kinetic process underlying dynamic disorder.
The ability of multidimensional spectroscopy to directly probe dark states, excited-state absorption features, and distinct coherent lineshapes between vibronic transitions poses challenges and opportunities to current models of polymer aggregates.
Detailed modeling of 2D coherent spectra is highly non-trivial due to the complexity of the full Holstein Hamiltonian needed to represent conjugated polymer systems. 
For example, it is likely necessary to consider models that are capable to account for many-body interactions in the nonstationary regime.~\cite{li_optical_2023-1} 
In such models, nontrivial line shift functions $g_1(t)$, with respect to population time and both coherence times, will remain in the optical response function, providing co-evolving complex but enriched lineshape features. 
Moreover, closely spaced vibronic states can also contribute to nonlinear lineshape evolution due to the nonadiabatic exciton-phonon coupling, which are further complicated due to inter and intrachain vibronic coupling with aggregation.~\cite{bittner_theory_2007, gutierrez-meza_frenkel_2021}

The ability to model these spectra will open new opportunities to understand the diversity of states involved in basic photophysical processes\cite{silva_efficient_2001,silva_exciton_2002}.
Such an approach will also help connect optoelectronic phenomena arising from many-body states, such as OLED PLQY roll-off at high currents due to bimolecular annihilation, to the structure of the material~\cite{stevens_exciton_2001}.
Moreover, this modeling will allow for better understanding of how the complex disordered energy landscape depends on electronic and vibrational couplings between or along chains, as dictated by the solid-state structure. 

\section*{Supplemental Material}
Additional information on the 2D coherent lineshapes and pulse characterization can be found in the supplemental material. 

\section*{Acknowledgments}
FT, HL, and CSA acknowledge funding from the Government of Canada (Canada Excellence Research Chair CERC-2022-00055). CSA acknowledges support from the Institut Courtois, Facult\'e des arts et des sciences, Universit\'e de Montr\'eal (Chaire de Recherche de l'Institut Courtois) and from the Natural Science and Engineering Research Council of Canada (Discovery Grant RGPIN-2024-05893). HK acknowledges funding from the National Science Foundation Graduate Research Fellowship (DGE-2039655). Work at Georgia Tech (HK, NS, and CSA) was supported by the Center for Soft PhotoElectroChemical Systems, an Energy Frontier Research Center funded by the DOE, Office of Science, BES under Award \# DE-SC0023411 (data analysis). ERB acknowledges funding from the National Science Foundation (CHE-2404788) and the Robert A.\ Welch Foundation (E-1337). MH acknowledges funding from the EPSRC (EP/T028513/1).

\section*{Author Declarations}
\subsection*{Conflict of Interest}
The authors have no conflicts to disclose. 

\subsection*{Author Contributions}
The following author contribution abides to the CRediT contribution taxonomy. 
Natalie Stingelin, Carlos Silva-Acu\~na, Eric R. Bittner, Hao Li and Félix Thouin were responsible for the conceptualization of the project. 
Data curation and formal analysis were performed by Félix Thouin, Elizabeth~Guti\'errez-Meza and Henry J.\ Kantrow.
Carlos-Acu\~na, Natalie Stingelin, and Eric R.\ Bittner were responsible for funding acquisition.
Henry J.\ Kantrow, Elizabeth~Guti\'errez-Meza, Hongmo Li, Félix Thouin, and Hao Li were involved in the investigation.
Félix Thouin and Carlos Silva-Acun\~a handled the project's administration.
The materials were synthesized by Martin Heeney and Qiao He; Hongmo Li was responsible for the polymer processing.
Software and algorithms were developed by Félix Thouin, Henry J.\ Kantrow, and Elizabeth~Guti\'errez-Meza.
Carlos Silva-Acun\~a and Natalie Stingelin were responsible for the supervision of students involved in the project.
Validation was performed by Henry J.\ Kantrow.
Visualization of data was handled by Henry J.\ Kantrow and Elizabeth~Guti\'errez-Meza. 
Henry Kantrow, Elizabeth~Guti\'errez-Meza, Félix Thouin, Hao Li and Carlos Silva were involved in writing the original draft.
All authors were involved in review and editing of the manuscript.

\subsection*{Data Availability}
Data and code used to produce the figures in this article are publicly available on the Borealis repository~\cite{kantrow_replication_2024}. 

\section*{Appendix: Experimental Methods}
\subsection*{Materials}
PBTTT-C16 was synthesized 
as previously reported ($M_n = 38$\,kg\,mol$^{-1}$, \DJ~$= 1.78$)~\cite{mcculloch_liquid-crystalline_2006}
and dissolved at 10\,mg\,mL$^{-1}$ in 1,2-dichlorobenzene (o-DCB, Sigma-Aldrich) at 85$^{\circ}$C for 1 hour.
Films were drop-cast on sapphire substrates preheated to $30^{\circ}$C and allowed to dry on the heated stage.

\subsection*{Coherent spectroscopy}
The coherent spectroscopy technique employed here has been described in previous publications~\cite{turner_invited_2011,thouin_stable_2018}. 
Briefly, the coherent optical beam recombination technique (COLBERT) uses a four-wave-mixing scheme to generate a third-order macroscopic coherent polarization, which is detected in amplitude and phase via spectral interferometry.
In the experiments reported below, we exclusively use a rephasing pulse sequence and phase matching, in which a conjugate beam pair interacts first, followed by another interaction after a delay called the population time ($t_{\mathrm{pop}}$).
The pulse fluence was 10\,nJ/cm$^2$ and the measured pulse duration was 18\,fs (Supplementary Figure S1).

\section*{References}

%




\begin{figure*}
\begin{center}
\noindent\Large{\textbf{Supplementary Material}
\vspace{0.2cm} 
\\\textbf{Quantum dynamics of photophysical aggregates in conjugated polymers}} \\
\vspace{0.5cm}

\normalsize
\noindent{\textit{Henry~J.~Kantrow, Elizabeth~Guti\'errez-Meza, Hongmo~Li, Qiao~He, Martin~Heeney, Natalie~Stingelin, Eric~R.~Bittner, Carlos Silva-Acu\~na, Hao~Li, and F\'elix Thouin}}\\
\end{center}
\end{figure*}

\renewcommand{\thefigure}{S\arabic{figure}}
\renewcommand{\thesection}{S\arabic{section}}
\renewcommand{\thesubsection}{\alph{subsection}}
\setcounter{figure}{0} 
\setcounter{section}{0} 

\begin{figure*}
\begin{center}
\includegraphics[width=10cm]{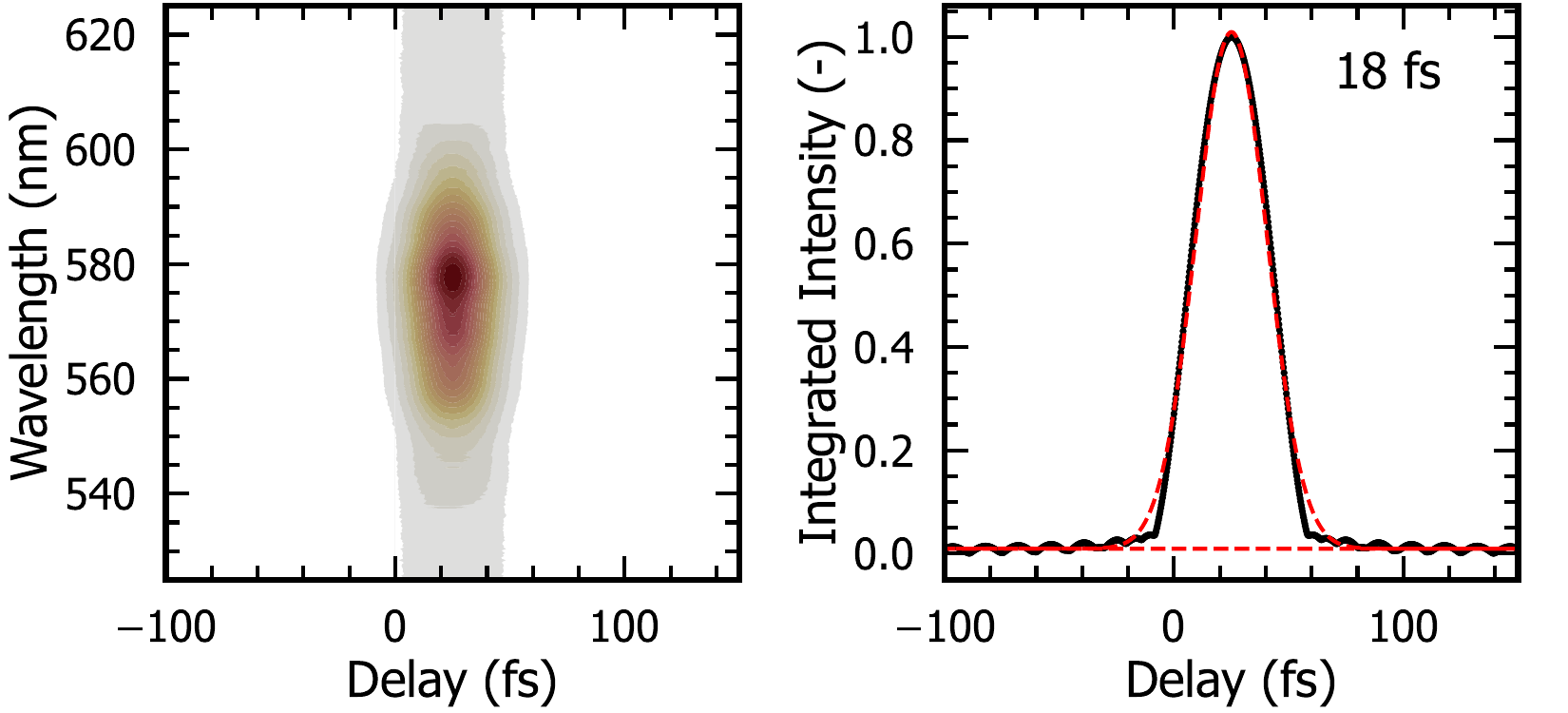}
\caption{SHG collinear FROG. Following the procedure of Amat-Rold\'an et. al.~\cite{amat-roldan_ultrashort_2004}, we estimate a pulse width of 18 fs. The pulse fluence was 10 nJ~cm$^{-2}$.}
\label{fig:xcorr}
\end{center}
\end{figure*}

\begin{figure*}
\begin{center}
\includegraphics[width=8.5cm]{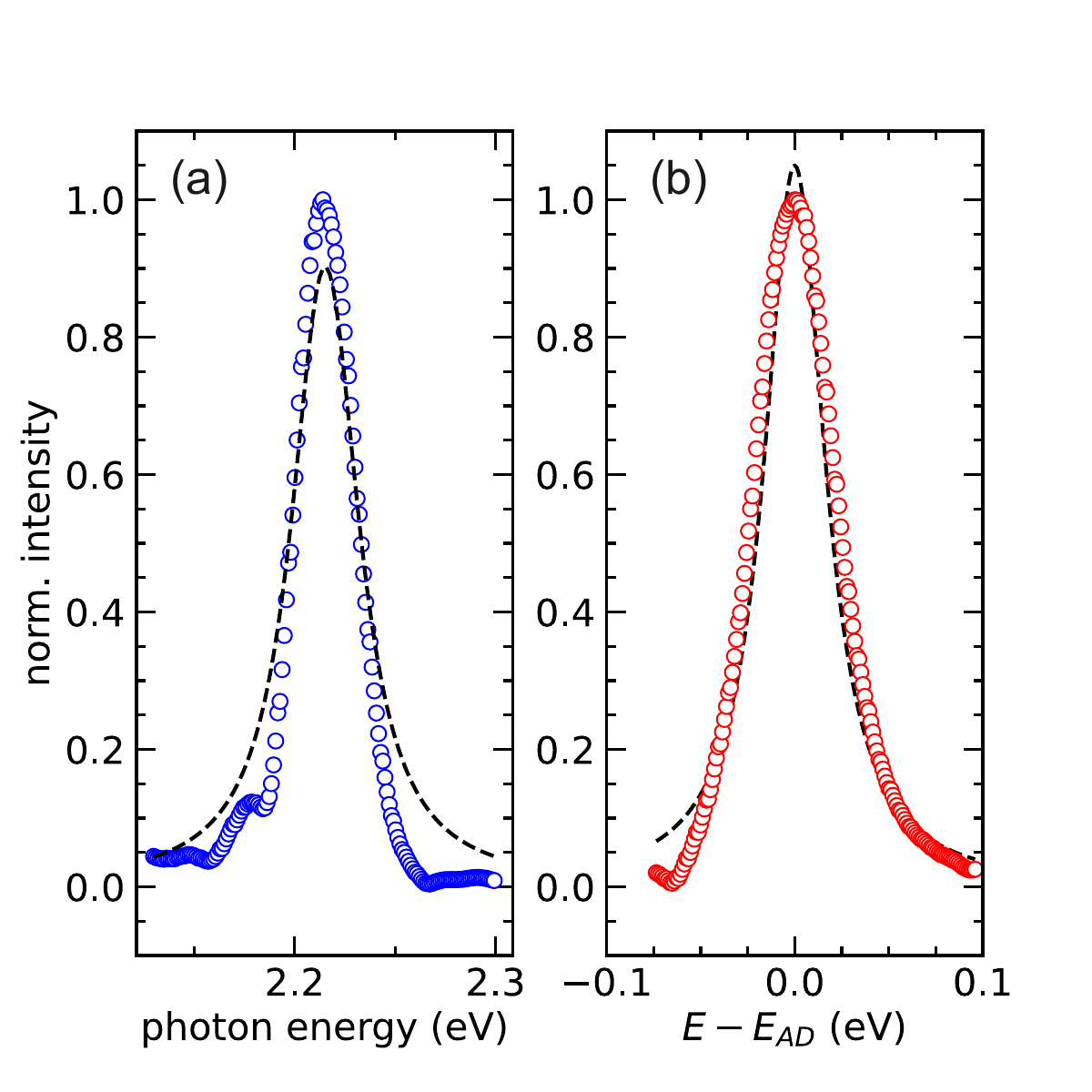}
\caption{(a) Diagonal and (b) anti-diagonal fit of the linewidth of the 0--1 vibronic transition at 2.21 eV. Because the homogeneous and inhomogeneous linewidths are of comparable magnitude, they must be fit simultaneously, as described elsewhere. \cite{siemens_resonance_2010} We estimate the homogeneous linewidth to be 37 $\pm$ 2\,meV and the inhomogeneous linewidth as 5 $\pm$\, 3 meV.}
\label{fig:fit01}
\end{center}
\end{figure*}

\begin{figure*}[t]
\begin{center}
\includegraphics[width=16cm]{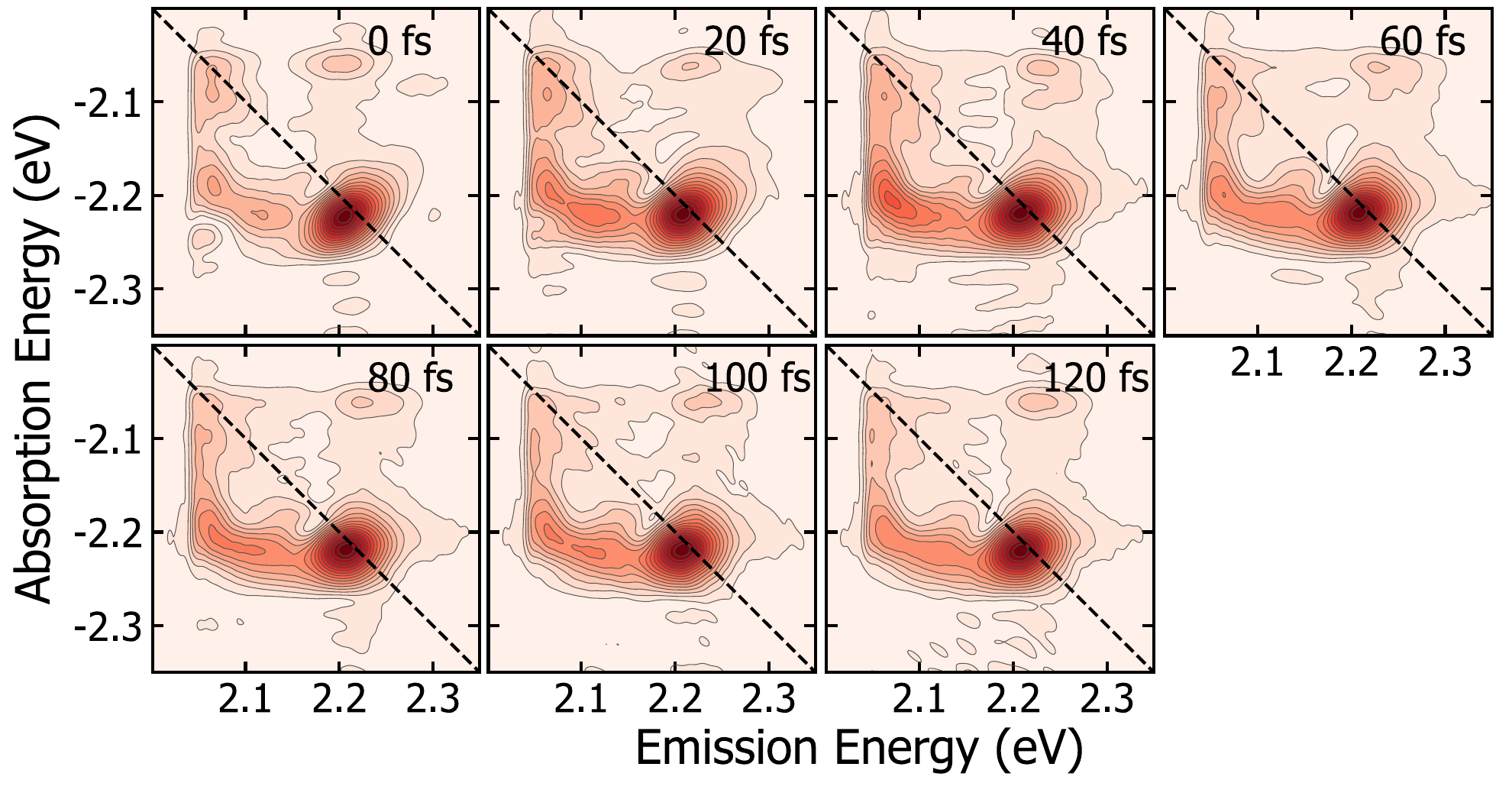}
\caption{Norm. of the 2D coherent spectrum using a rephasing pulse sequence at population times of 0 to 120 fs, in 20 fs intervals. Population times are labeled in the top right corner of each 2D map.}
\label{fig:norm_tpop}
\end{center}
\end{figure*}

\begin{figure*}
\begin{center}
\includegraphics[width=16cm]{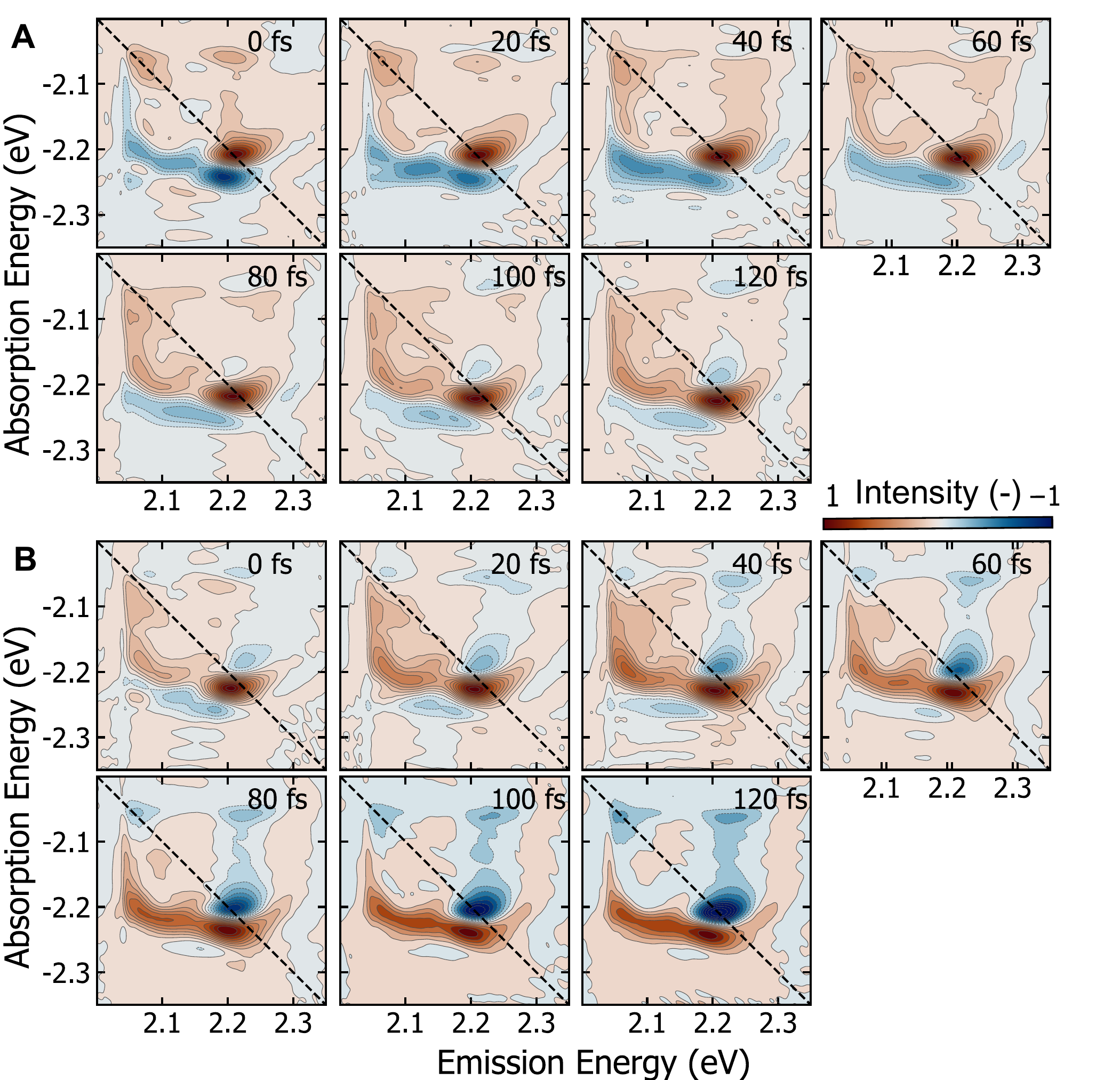}
\caption{(a) Real and (b) imaginary component of the 2D coherent spectrum using a rephasing pulse sequence at population times of 0 to 120 fs, in 20 fs intervals. Population times are labeled in the top right corner of each 2D map.}
\label{fig:norm_tpop}
\end{center}
\end{figure*}



\end{document}